\documentclass[
  ]
{article}

\usepackage{graphicx}
\usepackage{url}             

\begin{document}

\title{On Formation Of A Shock Wave In Front Of A Coronal Mass Ejection With Velocity Exceeding The Critical One
\thanks{Accepted for publication in Proceedings of the Solar Wind 12 Conference, Saint Malo, France, 21-26 June 2009}}


\author{M.V.~Eselevich \\
\it Institute of Solar-Terrestrial physics, Irkutsk, Russia }

\maketitle

\begin{abstract}
New study confirms conclusions made in \cite{Esel2008}; according
to it, there is a disturbed region expended along the CME
propagation direction in front of a coronal mass ejection whose
velocity $u$ is lower than the critical $u_C$ relative to the
surrounding coronal plasma. The time difference brightness (plasma
density) in the disturbed region smoothly decreases to larger
distances in front of CME. A shock wave forms at u higher than
$u_C$ in the front part of the disturbed region manifested as a
discontinuity in radial distributions of the difference
brightness.
\end{abstract}



\section{Introduction}

A coronal mass ejection (CME) structure in white light is often
characterized by the following well-known features: a bright
frontal structure (FS) that covers the region of decreased plasma
density (cavity) that may includes a bright interior (core).
However, besides the said features, another extended disturbed
region defined by \cite{Esel2007a} can exist immediately in front
of a CME. The aim of our study is to investigate changes in the
disturbed region form, when a CME velocity increases, and
possibilities for formation of a shock wave in this case.

\section{Method of analysis}

In the analysis, corona images obtained with the SOHO/LASCO C2 and
C3 \cite{Brue1995} were represented as the difference brightness
$\Delta P = P(t) - P(t_0)$, where $P(t_0)$ is the undisturbed
brightness at $t_0$ before the event considered, $P(t)$ is the
disturbed brightness at any instant $t > t_0$. Calibrated LASCO
images were used with the total brightness $P(t)$ expressed in
units of the mean solar brightness (P$_{msb}$).

Images of the difference brightness were employed to investigate
the CME dynamics and disturbed region. For the purpose we used
presentations in the form of isolines and sections both along the
solar radius at fixed position angles $PA$ and non-radial sections
at various instants $t$. On all the images, the position angle
$PA$ was calculated counterclockwise from the Sun's north pole.

\section{Data analysis}


First we consider two CMEs (CME1 and CME2) whose velocities $V$
differ greatly at $R =$ 4-5 R$_\odot$ (R$_\odot$ is the solar
radius). Two top panels of Figure~1 show the typical difference
brightness form (in isolines) for these two CMEs at the instants,
when their frontal structures FS appear in the C2 field of view.
Figure~1a presents the slow CME1 (5 May 1997, $V\approx 400$ km
s$^{-1}$), Figure~1c the fast CME2 (20 September 1997, $V\approx
800$ km s$^{-1}$). The values of velocity $V$ correspond to the
linear fit of the height-time measurements for the fastest frontal
parts of the CMEs and were taken from the CME catalogue
(\url{http://cdaw.gsfc.nasa.gov/CME_list/}).

\begin{figure}
\begin{center}
  \includegraphics[width=.8\textwidth]{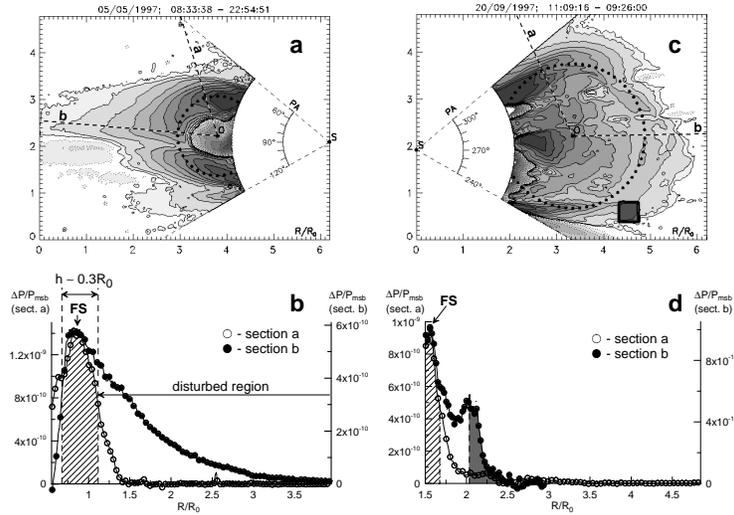}
\end{center}
  \caption{
  {\bf a, b} -- the slow CME1 on 5 May 1997;
  {\bf c, d} -- the fast CME2 on 20 September 1997;
{\bf top panels (a, c)} display images of the difference
brightness in the form of isolines; {\bf the bottom panels (b, d)}
present distributions of the difference brightness depending on
the distance $r$, measured from the frontal structure center
(point O) along two different sections ``a'' and ``b'', whose
directions are shown on the top panels.}
\end{figure}

Figures~1a,b show that on the images of both CMEs the frontal
structure FS can approximately be presented by a part of a circle
with its center at O (dots on the figures). The main direction of
the CME propagation that roughly coincides with its symmetry axis
is indicated by a heavy dashed line ``b''. It passes along the
streamer belt or streamer chains \cite{Esel1999,Esel2007b}; i.e.,
it is in the region of the quasistationary slow solar wind (SW).

In order to find the left boundary of the disturbed region (from
the CME side), by analogy with \cite{Mous1978} we determine the FS
width $h$ as a width at a half-height of the difference brightness
$\Delta P(r)$ distribution constructed from the FS center. For
CME1, the frontal structure in the direction of section ``a''
(Figure~1a) is least distorted by the disturbed region effect and
has a minimum width $h\approx 0.3$ R$_\odot$ (a curve with light
circles in Figure~1b). Let us take the FS right-hand boundary as
the left-hand boundary of the disturbed region, as shown in
Figure~1b. Its position is indicated by a vertical dashed line. In
Figure~1b, a curve with solid circles shows the $\Delta P(r)$
distribution along the section ``b'' in the direction of the CME
propagation. The disturbed region is marked by a horizontal line
with an arrow-head and labeled respectively.

The comparison between CME1 (slow) and CME2 (fast) yields two
principal distinctions:
\begin{enumerate}
\item The isolines that correspond to the minimum difference
brightness of the slow CME1 are extended along the direction of
its propagation, while those of the fast one are close in form to
a circle.
\item The difference brightness $\Delta P(r)$ distribution along the
direction of the CME propagation continuously decreases up to the
most specific front part of the disturbed region for the slow CME,
whereas in the front part of the fast CME disturbed region a
discontinuity appears in the $\Delta P(r)$ distribution on a scale
$\delta _F\approx$ 0.2-0.3 R$_\odot$ (shaded in Figure~1d).
\end{enumerate}

The $\Delta P(r)$ distributions constructed along the direction of
the CME propagation are presented for the set of eight CMEs in
Figure~2. CME velocities are different and increase from bottom to
top in Figure~2, thus the slowest CME with $V\approx 230$ km
s$^{-1}$ is shown on the bottom panel, and the fastest one with
$V\approx 2000$ km s$^{-1}$ on the top panel.

\begin{figure}
\begin{center}
  \includegraphics[width=.8\columnwidth]{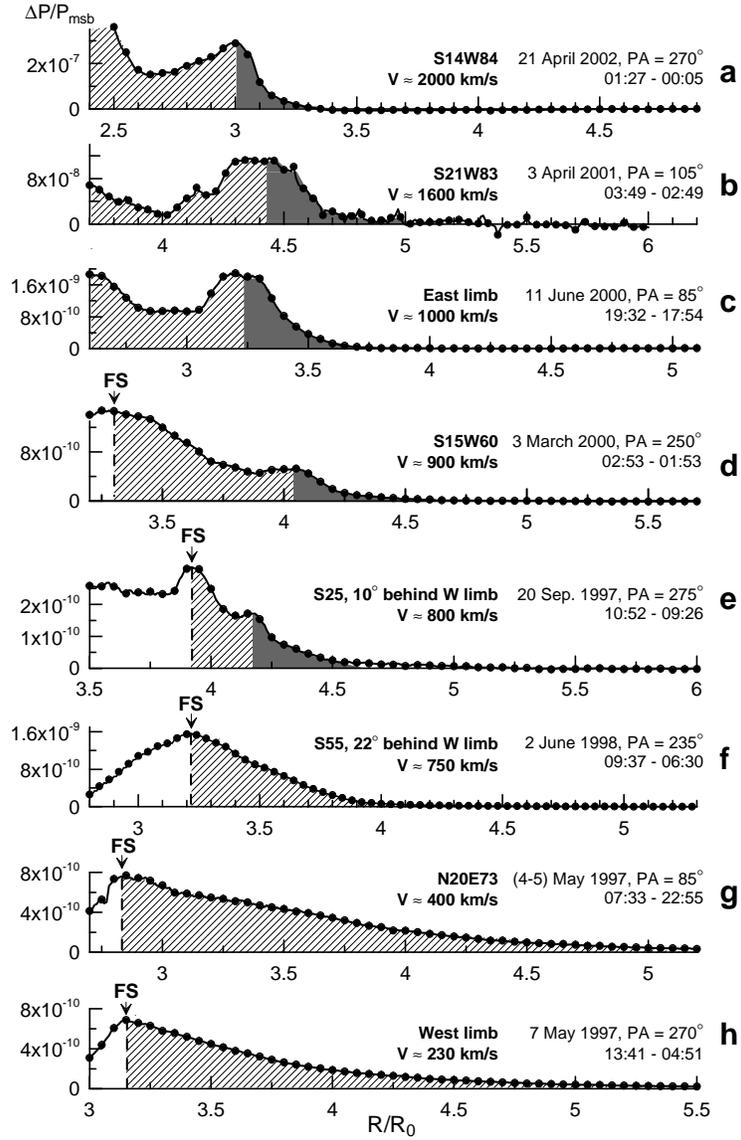}
\end{center}
  \caption{The $\Delta P(R)$ distributions for 8 CMEs along the direction of their propagation.}
\end{figure}

Slanted hatching in Figure~2 a-h presents the disturbed region and
half of the frontal structure (labeled as FS). As the CME velocity
increases from minimum ($V\approx 230$ km s$^{-1}$) to critical
one $V_C\approx$ 750-800 km s$^{-1}$, width of the hatched region
tends to decrease. Discontinuity in $\Delta P(R)$ distributions at
the front boundary of the disturbed region (shaded parts in
Figure~2 a-e) is observed when CME velocities $V$ are higher than
the critical velocity $V_C$. At the same time at $V < V_C$ such
discontinuity is absent, and the difference brightness
distribution smoothly decreases with increasing distance until it
becomes indistinguishable. Notice that CMEs with velocities $V$
close to the critical velocity VC (Figure~2 d-g) propagate near
the plane of the sky; thus, their measured velocity is close to
the real one (Figure~2 shows coordinates of places of the CME
initiation on the solar disk, taken from \cite{Crem2004} and {\it
Solar-Geophysical Data} (\url{http://sgd.ngdc.noaa.gov}).
Appearance of such a discontinuity at the front boundary of the
disturbed region at $V
> V_C$ was first described in \cite{Esel2008}.

Obviously, in processes of the ``CME -- undisturbed coronal
plasma'' interaction a crucial role should be played not by a $V$
value, but a value of CME velocity relative to the surrounding SW
stream $u = V - V_{SW}$.  Since CME velocities were determined in
the direction of their propagation, we took a velocity of the slow
SW flowing for the most part in the region of the coronal streamer
belt and streamer chains, along which the majority of CMEs move,
as the velocity $V_{SW}$ of the undisturbed solar wind
\cite{Hund1993}.

Figure~3 presents values of the relative velocity $u$ measured for
twenty four different CMEs at different distances. For $V_{SW}(R)$
we employed dependence derived by \cite{Wang2000} of the slow SW
velocity on the distance $R$ in the streamer belt. This dependence
is shown by a dash-dot line in Figure~3.

\begin{figure}
\begin{center}
  \includegraphics[width=.8\columnwidth]{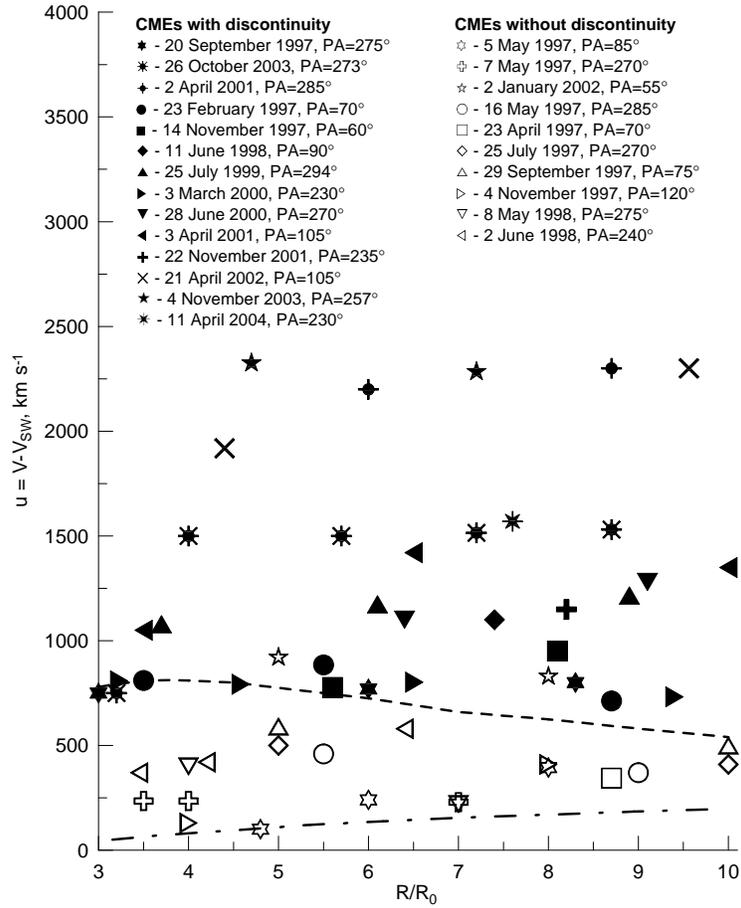}
\end{center}
  \caption{CME velocities $u = V - V_{SW}$ relative to the surrounding SW
  depending on a distance from the solar center for the CME without discontinuity (open marks)
  and the CME with discontinuity (solid marks). The dash-dot curve indicates the velocity $V_{SW}$
  of the quasistationary slow SW in the streamer belt from \cite{Wang2000}.
  The dash curve shows the Alfven velocity in the streamer belt from \cite{Mann1999}.}
\end{figure}

In Figure 3, solid marks correspond to the CMEs having a
discontinuity in the difference brightness distributions in front
of the disturbed region. CME velocities $V$ were determined from
the discontinuity motion. Open marks in Figure~3 indicate the CME
without discontinuity. In this case we took a velocity from the
CME catalogue (\url{http://cdaw.gsfc.nasa.gov/CME list/}).
Figure~3 shows that the cases with the discontinuity observed are
in the high-velocity region, and the cases without discontinuity
(the disturbed region smoothly decreased with distance is observed
there) are for the most part in the low-velocity region. Hence we
can assume that the discontinuity forms, when the relative CME
velocity $u$ exceed some critical $u_C$ value. A critical velocity
value may depend on a distance $R$.

Compare the obtained $u_C$ value with the typical velocity of
disturbance propagation in the magnetized corona plasma that is
roughly equal to the fast-mode MHD velocity $V_{MS}$ in the
plasma. In order to estimate $V_{MS}$ at these distances we may
use $V_A$ assuming that $V_{MS}\approx V_A$. Dashed line in
Figure~3 indicates the $V_A(R)$ dependence obtained by
\cite{Mann1999}.

Obviously in Figure~3 the Alfven velocity passes approximately
between clusters of points, which apply to the CMEs with
discontinuity and without it. Hence $u_C\approx V_A$, i.e., the
desired critical velocity is roughly equal to the typical velocity
of disturbance propagation in the magnetized plasma.

Hence we have a situation the classical gas dynamics refers to as
``transonic transition'' and formation of a shock wave. It was
predicted theoretically, but it is first observed experimentally
in the magnetized plasma.

\section{On possibility for resolution of a shock front width}

Here we will briefly discuss the problem of a possibility for
resolution of a shock front width in the corona. The discontinuity
is observed in distributions of the difference brightness $\Delta
P(R)$ that results from free-electron scattering and is averaged
along the line of sight in the optically thin corona. Since we do
not know exactly the matter-density distribution along the line of
sight, the observable scale $\delta _F$ in $\Delta P(R)$
distributions may differ from a real scale $\delta _N$ of the
plasma density discontinuity. As a result of the averaging the
observable discontinuity in the difference brightness profile can
have larger scale than the real discontinuity in the density
profile has.

In order to estimate an effect of such averaging, $\delta
_F/\delta _N$ ratios were found in the context of a simple
geometrical shock-front model in \cite{Esel2008}.
In the model considered, the
shock-wave front was represented as a spherical shell with an
outer radius $R_F$; the center of the shell was in the plane of
the sky at $R_C$ from the solar center (these parameters were
specified according to the CME form).

The brightness distribution $\Delta P(R)$, induced by
free-electron scattering within the shell in the range from the
shell center to its front edge, was calculated. At the given
distance $R$, the brightness value is defined by the integral
along the line of sight:
\begin{equation}
\Delta P(R) = \int\limits_{l} i(R,\theta )N(r)dl
\end{equation}
where $i(R,\theta )$ is the brightness induced by the one-electron
scattering, $N(r)$ -- density. The $i(R,\theta )$ function depends
on a impact distance $R$ and an angle $\theta$ relative to the
plane of the sky. In the spherical shell, the density was supposed
to change only depending on the distance $r$ from the sphere
center. In each case we chose a density profile $N(r)$ such that a
model brightness profile $\Delta P(R)$ obtained from $N(r)$ by
integration of Equation (1) showed the best correlation with the
experimental profile of the difference brightness $\Delta P(R)$.
Then a scale $\delta _N$ of the plasma density discontinuity was
found from the density profile $N(r)$ obtained.

Calculations show that the $\Delta P(R)$ profile broadening does
not exceed 20\% in comparison with the density profile $N(r)$.
Thus, scale of a brightness profile discontinuity is a good
approximation for determining scale of a density discontinuity in
a shock wave front.

It is worth noting that, at $R \geq 10$ R$_\odot$, a new
discontinuity (with thickness $\delta _F^* \ll \delta _F$) is
observed to form in the anterior part of the shock front for
events with $V > V_C$ considered above. Within the experimental
error, thickness $\delta _F^*\approx$ 0.1-0.2 R$_\odot$ does not
vary with distance and is determined by the LASCO C3 instrument
spatial resolution. Transfer from the shock front with thickness
$\delta _F$ to the discontinuity with thickness $\delta _F^* \ll
\delta _F$ can be interpreted as transition from the collisional
to collisionless shock wave \cite{Esel2009}. Similar
discontinuities in brightness profiles associated with
collisionless shock waves were registered ahead fast halo-type
CMEs ($V > 1500$ km s$^{-1}$) at distances $> 10$ R$_\odot$ in
\cite{Onti2009}.

\section{Conclusions}
It has been shown that in front of a coronal mass ejection having
a velocity $u$ lower than the critical $u_C$ relative to the
surrounding coronal plasma there is a disturbed region expended
along a direction of the CME propagation. The time difference
brightness $\Delta P$ in the disturbed region smoothly decreases
up to larger distances in front of the CME. Given $u > u_C$, a
discontinuity forms in distribution of difference brightness or
plasma density in the disturbed region front part. Since the $u_C$
value is close to the local fast-mode MHD velocity, which in
corona approximately equal to the Alfven one, the formation of
such a discontinuity when $u_C$ is exceeded may be identified with
the formation of a shock wave.




\subsection*{Acknowledgments} The work was supported by program
No. 16 part 3 of the Presidium of the Russian Academy of Sciences,
program of state support for leading scientific schools
NS-2258.2008.2, and the Russian Foundation for Basic Research
(Project No. 09-02-00165a). The SOHO/LASCO data used here are
produced by a consortium of the Naval Research Laboratory (USA),
Max-Planck-Institut fuer Aeronomie (Germany), Laboratoire
d'Astronomie (France), and the University of Birmingham (UK). SOHO
is a project of international cooperation between ESA and NASA.
The Mark 4 data are courtesy of the High Altitude
Observatory/NCAR.

\end{document}